\documentclass[12pt,a4paper, notitlepage]{article}

\usepackage{graphicx}
\usepackage{amsmath}
\usepackage{amssymb,amsfonts,latexsym}
\usepackage{bm}
\usepackage[mathcal]{euscript}
\usepackage{epsfig}
\usepackage{color}
\usepackage{authblk}

\renewcommand{\footnoterule}{%
  \kern -3pt
  \hrule width \columnwidth height 1pt
  \kern 2pt
}

\begin{document}

\title{The Physics of the Vicsek Model}

\author{Francesco Ginelli}
\affil{SUPA, ICSMB and Department of Physics, King's College, University of Aberdeen, Aberdeen AB24 3UE, United Kingdom}

\maketitle     

\begin{abstract}
In these lecture notes, prepared for the Microswimmers Summer School 2015 at
Forschungszentrum J\"ulich, I discuss the well known Vicsek model for
collective motion and its main properties.  In particular, I discuss 
its algorithmical implementation and the
basic properties of its universality class. I present results from numerical
simulations and insist on the role played by symmetries and
conservation laws. Analytical arguments are presented in an accessible and
simplified way, but ample references are given for more advanced
readings.
\end{abstract}


%
\section{Introduction}

Collective motion (or “flocking”) is a ubiquitous phenomenon, 
observed in a wide array of different living systems and on an even wider
range of scales, from fish schools \cite{Parrish} and mammal herds \cite{Ginelli2015} to bacteria
colonies \cite{Swinney2010} and cellular migrations \cite{Trepat},
down to the cooperative behavior of molecular motors and biopolymers 
at the subcellular level \cite{Schaller2010}. 
The aerial displays of starling flocks and
other social birds are of course among the most spectacular examples,
and have attracted the interest of speculative observers for quite a
long time \cite{Pliny}. 

To the physicist eye, these phenomena are also highly nontrivial
because they occur {\it far from equilibrium}, as single constituent
particles in a flock (whether they are birds, bacteria or cells) are {\it active},
i.e. they continuously
dissipate free energy to perform systematic (i.e. non-thermal) motion.
Also, collective motion often arises {\it spontaneously}, without any leader, external
field or geometrical constraint guiding the process. In a more
technical language, we may say that ordered motion follows from
the {\it spontaneous breaking of a continuous symmetry}, viewing
a collectively moving flock as an orientationally
ordered phase of {\it active matter} \cite{Ramaswamy2010}.

Collective motion phenomena, of course, are not restricted to living matter, and in recent times
they have been studied in various experimental systems, such as active
colloids \cite{Palacci} and driven granular matter \cite{Desaigne2010, Narayan}.  

The ubiquity of collective motion phenomena
at all scales, from groups of large vertebrates to subcellular
collective dynamics, strongly hints at the existence of some
universal features, possibly shared among the many different
situations, regardless of many individual-level details.
One way of approaching these problems is to construct
and study minimal models of collective motion, that is models
stripped of as many details as possible and only equipped 
with the basic features that we believe characterize the problem, typically its 
fundamental symmetries and conservation laws.
This approach is fundamentally justified by hydrodynamics
considerations, by which a great deal of microscopic details 
may be ignored, at least if we are interested in the large wavelength
and long time behavior of our system \cite{Forster}.

In any case, even if one is interested in finer, non-asymptotic details, it is surely
good practice, before starting toying with your favourite model, to
first understand the underlying, long wavelength physics inevitably
shared by all systems with the same fundamental features.

In these notes, I will introduce and discuss in details the properties of
the Vicsek model -- the simplest off-lattice model describing a flocking
state -- and of the related Vicsek class. Approaching the study of
collective motion, it is important to understand that all physical systems and models
sharing the same basic features with this class will also
display the same asymptotic properties. The only way to escape this, 
is to alter some fundamental property of the system, like changing the
broken symmetry (for instance from polar to nematic symmetry \footnote{Nematic
  systems are symmetric under a rotation by $\pi$.}) or
adding a further conservation law (for instance momentum conservation,
which is relevant for most active suspensions). 
This is likely going to be the main message of this lecture.

\section{The Vicsek model}

The Vicsek model (VM) is perhaps the simplest model displaying a transition
to collective motion; in the study of active matter plays a prototypical role,
similar to the one played by the Ising model for equilibrium
ferromagnetism. Its simple dynamical rule has been adopted as the starting point
for many generalizations and variations which have been applied to a
wide range of different problems.

The Vicsek model has been originally introduced 20 years ago by the
pioneering work of Vicsek and coworkers \cite{Vicsek95}. Subsequent
numerical studies (see for instance Refs. \cite{Gregoire2004} 
and \cite{Ginelli2008}) greatly helped in clarifying its properties.

\subsection{Definition and physical features}
\begin{figure*}[t]
\centering
\includegraphics[width=0.8\textwidth]{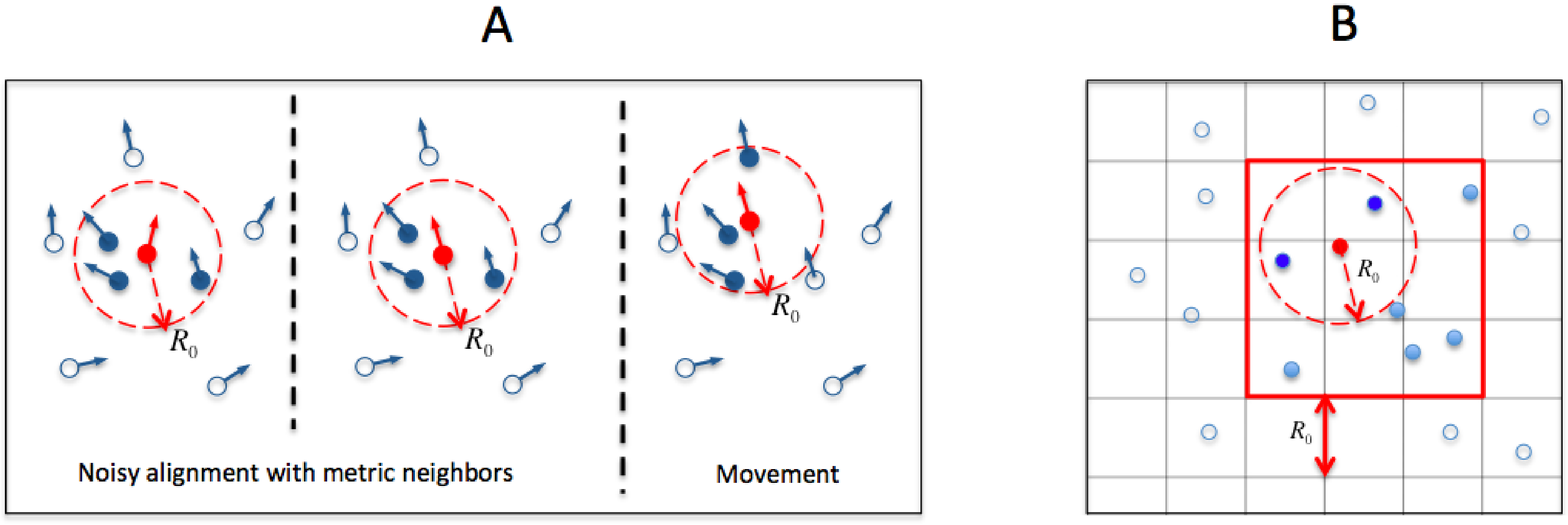}
\caption{(A) Cartoon depiction of the Vicsek dynamics. The particle in red
  aligns imperfectly with neigbours inside the metric range $R_0$ (dark
blue) and then moves along its heading direction. Note that neighbours may
change as a consequence of movement. In the cartoon only the red
particle adjusts its position and heading, but in the full algorithm all
particles update synchronously. (B) Molecular dynamics algorithm. Once
the system is divided in boxes of linear size $R_0$, we know immediately
that the candidate neighbouring particles of the red particle are
restricted to the ones laying in
the nine adjacent boxes inside the red square. These particles are marked
in light blue, but testing their distance from the red particle we
find that only some of them (dark blue) are actually closer than $R_0$
(note that even a particle laying in the same box as the red one may
occasionally be farther away than $R_0$).}
\label{fig:1}
\end{figure*}

The model describes the {\it overdamped} dynamics
of a collection of $N$ self-propelled particles (SPPs) characterized by their off-lattice position ${\bf r}_i^t$
and direction of motion (or heading) ${\bf s}_i^t$, a unit vector, $\left|{\bf s}_i^t\right|=1$. Here $i$ is the particle index,
$i=1,\ldots,N$, and $t$ labels time. 
All particles move with the same constant
speed $v_0$, according to the time-discrete dynamics
\begin{equation}
{\bf r}_i^{t+\Delta t} = {\bf r}_i^{t} + \Delta t \, v_0 {\bf s}_i^{t+\Delta t}
\label{eq:1}
\end{equation}
so that orientation ${\bf s}$ and particle velocity ${\bf v}=v_0 {\bf s}$ coincide
but for a multiplicative constant (and often the term {\it velocity}
is also used for the orientation {\bf s}).\\
Particles tend to align their direction of motion with the one of
their {\it local} neighbours, and ${\bf s}_i^t$ depends on the average direction of
all particles ($i$ included) in the spherical neighborhood
$\mathcal{S}_i$ of radius $R_0$ centered on $i$. Indeed, in the Vicsek
algorithm the alignment with ones neighborhood is almost perfect, only
hampered by a white noise term which plays a role analogous to the one
of a temperature in equilibrium systems. In two spatial dimensions ($d=2$), 
the direction of motion is defined by a single angle $\theta_i^t$,
with ${\bf s}=(\cos \theta, \sin\theta)$, and
one may simply write the orientation dynamics as 
\begin{equation}
\theta_i^{t+\Delta t} = \mbox{Arg}\left[ \sum_j n^t_{ij}
  {\bf s}_j^t \right] + \eta \, \xi_i^t
\label{eq:2}
\end{equation}
where $\xi_i^t$ is a zero average, delta-correlated scalar noise 
\begin{equation}
\langle \xi_i^t \rangle =0 \;\;\;\;,\;\;\; \langle \xi_i^t \,\xi_j^k
\rangle \sim \delta_{tk}\delta_{ij}
\label{eq:noise}
\end{equation}
uniformely distributed in $\left[ -\pi,\pi\right]$\footnote{With this
  choice, $\eta=1$ is the largest meaningful noise amplitude. At
  each time step, it completely randomizes all particle headings in the interval $[-\pi,
  \pi]$. Thus the case $\eta=1$ completely dominates alignment and just gives a
  collection of independent random walkers.}. Such a noise is
often called {\it white}, since it has a flat Fourier spectrum.\\
In Eq. (\ref{eq:2}), the function Arg returns the angle defining the
orientation of the average vector $\sum_j n^t_{ij}
  {\bf s}_j^t$, and $n^t_{ij}$ is the connectivity matrix,
\begin{equation}
n^t_{ij} = \left \{
\begin{array}{c}
1\;\;\;\;\mbox{if} \left|{\bf r}_i^t - {\bf r}_j^t\right|<R_0\\
0\;\;\;\;\mbox{if} \left|{\bf r}_i^t - {\bf r}_j^t\right|>R_0\\
\end{array}\right.
\label{eq:metric}
\end{equation}
This way of chosing neighbours is sometimes defined as {\it metric},
being based on the metric notion of distance. 
The dynamics (\ref{eq:1})-(\ref{eq:2}), depicted in Fig.~\ref{fig:1}A, is {\it synchronous}, meaning that
all particles positions and headings are adjusted at the same time.\\
In studying this model, one can always chose a convenient set of space
and time units, such that $\Delta t = R_0 =1$ and the model behavior only
depends on three {\it control parameters}: the noise amplitude $\eta$, the particles
speed $v_0$ and the total density of particles $\rho_0 = N/V$, where $V$ is
the volume of the system. 
Being interested in the bulk
properties of a system, one typically assumes periodic boundary
conditions, so that $V=L^d$, with $L$ being the linear system
size. In numerical simulations, periodic boundary conditions help to minimize finite size
effects due to finite boundaries, and in the following we will
implicitly assume them unless stated otherwise.

In the literature, one may find a number of slightly different
flavours of the algorithm defined above. For instance, the noise in
Eq. (\ref{eq:2}) may be distributed according to a Gaussian, a small,
short ranged repulsion force between particles may be included to
account for volume exclusion, or the
position ${\bf r}_i$ at time 
$t+\Delta t$ , as defined in Eq. (\ref{eq:1}), may be determined by the direction of motion at time $t$
and not at time $t+\Delta t $ (indeed, this was
  actually the choice made in the original paper by Vicsek and
  coworkers \cite{Vicsek95}). However, typically
all these differences do not matter much, and do not change the physical
properties of the Vicsek model. \\
On the other hand, there are some features
which are essential, and define what we call the {\it
  Vicsek class}. It is worth discussing them explicitly:
\begin{itemize}
\item
{\it Spontaneous symmetry breaking to polar order}. Eqs. (\ref{eq:1})-(\ref{eq:2})
are isotropic in space, as no preferred direction is given
a priori. However, Eq. (\ref{eq:2}) contains an explicit polar (or
ferromagnetic) alignment
term. If this alignment term is strong enough to overcome the effect
of the noise (or to put it differently, if the noise amplitude $\eta$
is low enough), the system may develop global orientational order and
thus collective motion, signaled by a
finite {\it polar order
  parameter} (or center of mass velocity)
\begin{equation}
{\bm \varphi}(t) = \frac{1}{N}\sum_{i=1}^N {\bf s}_i^t 
\label{op}
\end{equation}
an analogous of the total magnetization in spin systems.
The value of the modulo of the polar order parameter $\varphi(t)=|{\bm
  \varphi}(t) |$ is essentially determined (minus fluctuations and
finite size effects) by the three
control parameters $\rho_0$, $\eta$ and $v_0$. Its stationary time
average $\phi = \langle \varphi(t) \rangle_t$ is typically used to
describe the spontaneous symmetry breaking phenomenon, with $\phi > 0$
in the ordered phase\footnote{There are of course finite size effects,
  and in the disordered phase the vectors ${\bf s}_i$ do not
  cancel exactly one with each other, leading to
  $|{\bm \phi}| \sim \frac{1}{\sqrt{N}}$.}. 

Its orientation in the ordered phase, on the other hand, is not determined
a priori, and all directions are equally likely 
(the one picked up at a given time being chosen by fluctuations).
Since the orientation can change continuously in space\footnote{As opposed,
for instance, to the Ising model in which spins may only assume two
values, $\pm 1$.}, in the transition to collective motion 
a {\it continuous} symmetry is spontaneously broken. This has a number of
important consequences that will be explored in this notes.

\item {\it Self-propulsion and local alignment interactions}.
Particles are self-propelled, that is, they move according to
Eq. (\ref{eq:1}). In particular, they change their relative position
according to their velocity fluctuations $\delta {\bf s}_i^t={\bf
  s}_i^t - {\bm \varphi}(t)$. Thus, the connectivity matrix in
Eq. (\ref{eq:2}) is not static, but it changes in time in a nontrivial
way. This is exactly where the far from equilibrium nature manifests 
in the Vicsek model, as it will be discussed in section \ref{MW}.
Of course, the connectivity matrix will change as a consequence of
particle motion only if the interactions are local, that is, if $R_0
<< L$.

\item {\it Conservation laws}.
The only conservation law of the VM is the conservation of
the total number of particles, that is, our birds do not die or get
born on the fly. There are no other conservation laws, and in
particular it should be noted that momentum is not conserved. Our
self-propelled particles are thought to be moving over a dissipative
substrate (or in a viscous medium) which acts as a momentum sink. 
This is of course not the case of a particle swimming in a
three-dimensional suspensions \footnote{In quasi-two dimensional suspensions
momentum can be dissipated at the boundaries.}, where momentum is
transferred from the swimmers (typically exerted as a force dipole) 
to the surrounding fluid, and long-range hydrodynamic
interactions are probably relevant (and for man-made
micro swimmers or self-propelled nano-rods they are typically the
only interactions!).

As a consequence of the lack of momentum conservation, also Galileian
invariance is broken. In fact, the VM is explicitly formulated in the
reference frame in which the
dissipative substrate is at rest, and it is not invariant under any arbitrary
velocity shift. 
\end{itemize}
All together, the features discussed above  define the Vicsek
class. Finally, we have to remark another obvious feature of the Vicsek model:
all particles move with the same speed $v_0$. However, to
a certain extent it is possible to relax this conditions staying
inside the Vicsek class. For instance, one can let
the individual speeds fluctuate in some bounded interval without
changing the model asymptotic properties.

\subsection{Vectorial noise and Vicsek model in three space dimensions}

In the literature, it is possible to find different ways of implementing
the noise in the equation for the orientation dynamics. In the past,
some attention has been given to the so called {\it vectorial} noise \cite{Gregoire2004} 
(as opposed to the noise used in Eq. (\ref{eq:2}) which is sometimes
called {\it scalar} or {\it angular}). One may replace Eq. (\ref{eq:2}) by
\begin{equation}
{\bf s}_i^{t + \Delta t} = \frac{\sum_j n_{ij}^t {\bf s}_j^t + \eta \,
  m_i {\bm \xi}_i^t}{\left\| \sum_j n_{ij}^t {\bf s}_j^t + \eta \,
  m_i {\bm \xi}_i^t \right\|}
\label{eq:2bis}
\end{equation} 
where $m_i = \sum_j n_{ij}$ is the number of interacting neighbours,
and ${\bm \xi}$ is a random unit vector, delta-correlated in time and
in the particle index. The denominator in the r.h.s. of
Eq. (\ref{eq:2bis}) is a normalization term to ensure that $|{\bf
  s}|=1$.
If one interprets the scalar noise of
Eq. (\ref{eq:2}) as an error the SPP makes trying to take the
(perfectly determined) mean direction of motions of his neighbours, the vectorial version of the
noise can be thought as the sum of the errors made while trying to assess the
direction of motion of the interacting neighbours\footnote{In the light of the
  central limit theorem, a noise prefactor proportional to $\sqrt{m_i}$
  would be more appropriate than one directly proportional to $m_i$,
  but here I will stick to the latter mainly for historical
  reasons.}. Certain literature, also refers to these two noises
implementation as (respectively) intrinsic and extrinsic, but it is
important to stress that {\it these two implementations do not yield
different asymptotic properties}, even if their finite size behavior
may be slightly different (more later on this). 

It is however worth remarking that Eq. (\ref{eq:2bis}) can be
directly extended to any spatial dimension, while starting from Eq. (\ref{eq:2})
requires some more care. In order to write a Vicsek dynamics with scalar noise in
$d=3$, one has to introduce a rotation operator ${\mathcal R}_{\eta}$
performing a random (and of course delta-correlated) rotation uniformly distributed around the
argument vector, 
\begin{equation}
{\bf s}_i^{t + \Delta t} = {\mathcal R}_{\eta}\left[ \frac{\sum_j n_{ij}^t {\bf s}_j^t}{\left\| \sum_j n_{ij}^t {\bf s}_j^t \right\|}\right]
\label{eq:2-3D}
\end{equation} 
In $d=3$, for instance, ${\mathcal R}_{\eta}[{\bf
  s}]$ will lay in the solid angle subtended by a spherical cap of amplitude
$4\pi\eta$ and centered around ${\bf s}$. 

\subsection{Limiting cases}

It is instructing to consider the relations between the Vicsek model
and some well-known models of equilibrium statistical physics.

Obviously, the VM may be seen as an XY (or Heisenberg
in $d=3$) ferromagnet in which particles are not fixed in some
lattice positions but can actually move along the spin
direction. Indeed, the XY or Heisenberg equilibrium models can be
formally recovered in the case $v_0 \to 0$, where particles
do not move at all and $n_{ij}$ is fixed once for all. 
If the local connectivity of the static connection
network its dense enough, its dynamics converges to the equilibrium distribution 
of an XY or Heisenberg model, with a temperature $T$ that is
a monotonic function of the noise amplitude $\eta$. 
This is however a  {\it singular limit}\footnote{Singular means that the $v_0=0$ case is
radically and qualitatively different from the behavior at any small
but finite $v_0$.}. 

Another way of looking at the Vicsek model is to see it as a {\it
  persistent random walk} in which particles may align their
directions of motion one with each other by some local interaction
rule. 
In continuous time and $d=2$, the persistent
random walk can be written as 
\begin{equation}
\dot{\bf r}_i = v_0 {\bf s}_i\;\;\;\;,\;\;\; \dot{\theta}_i = \xi_i
\label{eq:pers}
\end{equation}
(with $\xi_i$ being some white noise) which is just a Vicsek dynamics without the alignment interaction term
(starting from Vicsek dynamics, Eq. (\ref{eq:pers}) can be formally
obtained by taking the limit $R_0 \to 0$).
Once again, this is a singular limit, and a collection of
non-interacting persistent random walker has an equilibrium
distribution with some temperature given by the noise term. 

The opposite limits, $v_0 \to \infty$ and $R_0 \to \infty$, also
correspond to singular cases. As already mentioned, if the
interactions are long ranged, the system is globally coupled and
the connectivity matrix $n_{ij}$ is trivially static. In this way,
motion is completely decoupled from long-ranged alignment, and
most (if not all) of the fascinating Vicsek model properties are lost.

The infinite speed limit $v_0 \to \infty$, on the other hand, just
produces a random rewiring of the connectivity network: if $v_0 >> L$,
any small fluctuation $\delta {\bf s}_i^t$ in the orientation will push 
nearby particles infinitely apart. In a system with periodic boundary conditions this
is equivalent to random rewiring of interactions, another trivial case in which
motion decouples from alignment.

The bottom line is that, while is interesting to understand the
relations between the VM and its limiting cases, it is not
possible in general to deduce properties of the former from the study
of the latter (singular) limiting cases\cite{Ginelli07}.

\subsection{Algorithmic implementation}

The Vicsek model is extremely simple and particularly well suited for
numerical studies, as Eqs. (\ref{eq:1})-(\ref{eq:2}) can be
easily implemented on a computer. However, it should be noted that
a straightforward implementation of the metric neighbouring
condition (\ref{eq:metric}) would require testing the distance of all
$i-j$ couples, an operation scaling with system size as order $N^2$. This approach
would quickly become unmanageable  as the number of SPPs $N$ grows,
making practically impossible to run simulations with more than a few
thousands particles.

There is of course a way around this problem, based on techniques
originally developed for the study of molecular dynamics. The idea is
rather simple, even if its algorithmic interpretation may not be so
straightforward. One should ideally divide the system volume
$L^d$ in boxes of linear size $R_0$ (remember that one can always
rescale space so that $R_0=1$), assigning at each timestep each
particle to a given box. Once this is done, it is clear that for any
given particle $i$, all other particles laying outside the box
containing $i$ and its next neighbouring boxes cannot be closer than
$R_0$. Therefore, one immediately and effortlessly reduces its search
to a handful of boxes per particle. In $d=2$ one has to only look into 
9 boxes (the general formula in spatial dimension $d$ of course gives
$3^d$ boxes). A sketch for this algorithm is depicted in Fig.~\ref{fig:1}B.

At any fixed total density \footnote{Finite size analysis is performed increasing both the
  number of particles $N$ and the total volume $V=L^d$ in such a way
  that the total density $\rho_0=N/V$ stays constant.}, the mean number of
particles contained in these boxes does not grow with
$N$, so that the
number of operation needed to find all the interacting couples grows
only linearly with $N$. Since also assigning particles to boxes is an
order $N$ operation, it is immediate to conclude that the entire
molecular dynamics algorithm computational time is of order $N$ rather
than $N^2$ as the naive algorithm. A huge improvement if one is
interested in asymptotic (i.e. long time and large $N$) properties. 

Any serious numerical study should employ molecular dynamics
algorithms. Current state of the art simulations of Vicsek model
involve from a few millions to a few tens of millions of particles.


\section{Physical properties}

We now proceed to discuss the main physical properties exhibited by the
Vicsek class. As we shall see, they mostly emerge from the intriguing
interplay between particles self propulsion and the spontaneous
symmetry breaking characterizing the ordered state.

\subsection{Transition to collective motion and phase separation}

Numerical simulations easily show that the Vicsek model display a
transition from disorder to ordered collective motion. 
For instance, as the noise amplitude $\eta$ is decreased below a certain threshold (and both $\rho_0$ and
$v_0$ kept fixed), particles start to synchronize their heading and to
move together. Starting from disordered initial condistions, this
coarsening process is relatively fast, and the size $\ell_d$ of
ordered domains grows linearly in time, $\ell_d \sim t$ \cite{Ginelli2008}.

The easiest way to capture the transition to collective
motion is to monitor the order parameter ${\bm \varphi}$ (the center
of mass velocity) defined in
Eq. (\ref{op}). At high noise amplitudes, SPPs are unable to synchronize
their headings, which tend to cancel out in the sum $\sum_i {\bf
  s}_i$. It can be shown that the sum of $N$ randomly oriented unit
vectors has a modulo of order $\sqrt{N}$, so that in the disordered
phase the scalar order parameter $\varphi \sim \frac{1}{\sqrt{N}}$, or
essentially zero for any large number $N$ of SPPs.

At lower noise amplitudes, below a certain threshold $\eta_c$, the system undergoes a
spontanous symmetry breaking phase transition as SPPs synchronize their heading. The
scalar order parameter becomes 
finite and roughly of order one (note that perfect
order exactly implies $\varphi=1$).

\begin{figure*}[t]
\centering
\includegraphics[width=0.53\textwidth]{fig2.eps}
\includegraphics[width=0.27\textwidth]{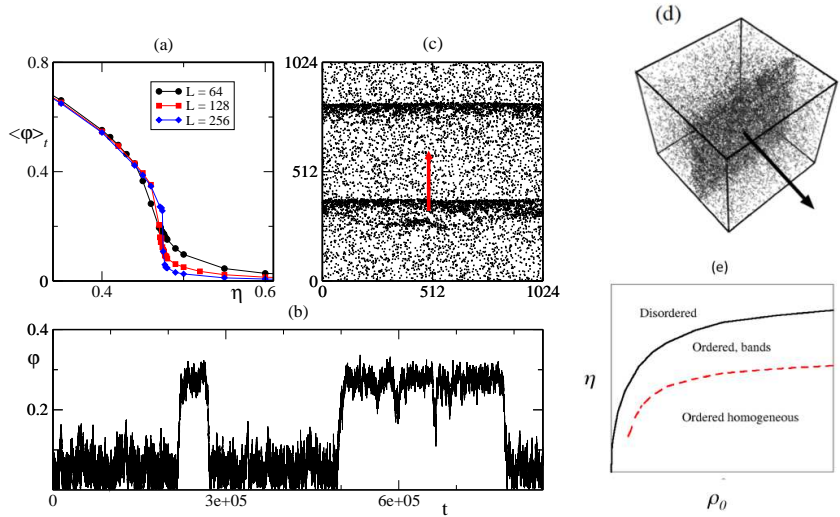}
\caption{(a) Characteristic order parameter curves vs. noise amplitude
  for fixed density and different system sizes (see legend). (b) Typical scalar order parameter timeseries at the onset
  of order, showing bi-stability between order and disorder.  (c)
  Ordered bands in $d=2$.  (d) Ordered sheet in $d=3$.  (e) Qualitative VM phase
  diagram in the $(\rho_0,\eta)$ plane. The arrows in panels (c)-(d) show the direction of
  motion. For more details, see reference \cite{Ginelli2008}.}
\label{fig:2}
\end{figure*}

This is resumed in Fig.~\ref{fig:2}a, where the long-time (or
stationary) average $\phi=\langle \varphi(t) \rangle_t$
is shown for different noise amplitudes. The parameter that is varied
as the system goes through the symmetry breaking is referred to as
{\it control parameter}. 
The threshold noise amplitude value for the onset of
collective motion is, of course, not independent from the other model
parameters, and one has $\eta_c=\eta_c(\rho_0, v_0)$. 

One simple way to understand the onset of collective motion is to consider
that, in order to synchronize the heading of all SPPs, information
should be able to propagate through the entire system. While
alignment interactions between particles produce such information,
noise clearly destroys it. A simple mean-field like argument can then
be put forward for low densities. To simplify things, lets
rescale our units so that the interaction range is one, $R_0=1$.
If $\rho_0 << 1$ particles are often isolated, and their relatively rare
interactions can be treated as {\it instantaneous} collisions\footnote{To be
  more precise, we want the mean inter-collision time to be much
  larger than the time two nearby particles spend at a mutual distance
  shorter than $R_0$.} from which particles emerge agreeing on their
headings. The distance $\ell$ that a particle travels between collisions,
the {\it mean free path}, scales as $\ell \sim 1/\rho_0$. Information
can propagate through the system only if the mean free path is larger
than the SPP {\it persistence length} $\ell_p$, that is the distance a
particle can travel before losing its out-of-collision heading. 
At the onset of order one expects these two quantities to have the
same magnitude, $\ell \sim \ell_p$. Given that the persistence length
is inversely proportional to noise variance, $\ell_p \sim
v_0/\eta^2$, we have immediately 
\begin{equation}
\eta_c \sim \sqrt{\rho_0}
\label{mf}
\end{equation}
a relation that has been numerically verified for $\rho_0 << 1$
(at least in $d=2$), and that  defines a critical line in the
$(\eta,\,\rho_0)$ plane. This implies that one can also use the total
density as a control parameter, keeping the noise amplitude fixed. In
this case, one crosses to collective motion as the density is increased.

At a first glance, one may think that the symmetry breaking transition
to collective motion should be similar to the transition to order in
an equilibrium spin system, leading directly to some homogeneously ordered 
state. This is however not the case, due to the interplay between
local order and local density induced by motion. Moving particles,
indeed, may gather in high density patches, increasing in turn the
number of interacting neighbors, i.e. particles with a mutual distance
smaller than $R_0$. Locally high density has a positive feedback on
the efficiency of the alignment interaction, so that high density
patches may be able to locally align while the rest of the systems
does not; this is something that cannot happen in an equilibrium spin
system!

One can indeed show that this feedback mechanism inevitably leads
to a long wavelength instability near the onset of order \cite{Bertin1, Bertin2},
that destabilizes the homogeneous ordered phase and leads to (spontaneous) phase
separation. For the polar symmetry of the Vicsek class, these phase
separation takes the form of high-density ordered bands that travel
in a low-density sea of disordered particles \cite{Gregoire2004,
  Ginelli2008} (see
Fig.~\ref{fig:2}c). 
Bands extend transversally to the direction of motion and are
characterized by a well-defined width, so that it is possible to
accomodate several in the same system. Indeed, on very large timescales
they seem to settle in a regularly spaced pattern, leading to a
smectic arrangement of traveling ordered bands \cite{Tailleur2}. 
In $d=3$, simple symmetry considerations imply that these structures
manifest as sheets, again extending transversally to the direction of
motion (Fig.~\ref{fig:2}d). 

At lower noise values or larger densities, the long wavelength instability disappears,
and a second transition leads to a homogeneous ordered phase. The
resulting Vicsek class phase diagram, sketched qualitatively in 
Fig.~\ref{fig:2}e, is thus composed of three phases. A disordered one,
akin to a collection of persistent random walkers, a phase-separated ordered
regime, characterized by high density ordered bands, and finally an
homogeneous ordered phase. In the latter two phases, the rotational
symmetry is spontaneously broken and the system exhibit collective motion.
 
As a consequence of phase separation, the symmetry breaking transition
to polar order is a first order one, rather than a second order
critical one \cite{Gregoire2004, Ginelli2008}. At the onset of order, the system is bistable, and
alternates between the disordered regime and the appearance of a
single ordered band, with the order parameter $\varphi(t)$ showing the
corresponding jumps characteristics of phase coexistence and of first
order phase transtions (Fig.~\ref{fig:2}b). 

The transition to collective motion can also be interpreted as a 
liquid-gas transition, albeit in a non-equilibrium context and with
no accessible supercritical region \cite{Tailleur2}.

This phase diagram, with phase separation and a first order transition
characterizing the onset of order, is rather generic; it is indeed common not only to
the entire Vicsek class, but also to systems whose broken state is
characterized by a different symmetry (although details of the phase
separated regime may change with different symmetries). However, the existence of this
phase separated regime has proven rather elusive, and it took a decade 
from the first introduction of the Vicsek model to discover it. 
In fact, the long-wavelength instability leading to phase separation
is characterized by a rather large instability wavelength
$\Lambda_c$, so that in systems not too large, where $L<\Lambda_c$, phase
separation cannot be observed and the transition may be mistakenly
thought to be continuous
and critical. It is only when $L$ is sufficiently larger than
$\Lambda_c$ that the true asymptotic behavior of the Vicsek model
emerge.

The instability wavelength -- of course -- depends on
model parameters and, to make things worse, also on non-universal
details such as the noise implementation. In particular, it is rather
larger in systems with a scalar noise (as in Eq. \ref{eq:2}) than in systems with a vectorial
one (as in Eq. \ref{eq:2bis}), so that it is not uncommon to be able
to observe bands only in systems with several hundred thousands of
particles or more. Moreover, $\Lambda_c$ seem to diverge both in the
low density and in the low speed limits \cite{Ginelli2008}.
These difficulties (one can say that the VM is characterized
by very strong finite size effects) have fueled a long
debate on the order of the phase transition, on the genericity of the
phase separated band regime and on the eventual difference between
scalar and vectorial noise models in the thermodynamic limit. 
Careful finite size analisyis and large scale simulations, together
with the study of hydrodynamic theories \cite{Bertin1, Bertin2} for the Vicsek class however,
have gathered convincing evidence over the last decade. Nowadays there is a general
consensus for the scenario detailed above: no asymptotic difference
between scalar and vectorial models, first order transition to
collective motion and genericity of the phase separation
scenario.\footnote{The stability of bands may depend on the nature of
  the boundaries. They are much favoured in periodic boundary
  conditions but one may think of other boundaries which frustrate
  them, for instance reflecting circular boundaries. 
Nevertheless, simulations with frustrating boundaries show that travelling bands emerge in the
bulk of the system and travel up to the frustrating boundaries, where
they disintegrate, thus further supporting their genericity.} 

We conclude this section noting that moving bands quite similar to VM ones have been
observed in {\it in vitro} experiments with motility assays, i.e. in a mixture
of molecular motors and actin filaments which are among the
constituents of cellular cytoskeleton \cite{Schaller2010}. Such a systems is of course
much more complicated than the VM, but is still characterized by
self-propulsion (due to the molecular motors) and may undergo a
spontaneous symmetry breaking thanks to filament interactions which
are effectively aligning. These experimental result demonstrate the power of the
minimal model approach.

\subsection{Topological Vicsek model}
\label{topo}

A relevant change of the Vicsek rule (\ref{eq:2}) is given by
topological interactions\cite{Ginelli2010b}. In topological models, one choses interacting neighbours not as
the SPPs lying inside a metric range $R_0$, but on the basis of some
local topological (or metric-free) rule, such as the $n_c$ nearest neighbours or the
Voronoi neighbours\footnote{In the Voronoi algorithm, at each timestep
  one constructs a Voronoi tessellation centered on particle
  positions. The interacting neigbours of a particle $i$ are then chosen
  to be the $j$ particles forming the first shell around particle $i$
  in the Voronoi tessellation. To simulate this algorithm, one cannot
  use the molecular dynamics techniques of metric models, but should 
resort to libraries optimized for geometric tessellations. A good (and
freely available) example is the CGAL library, http://www.cgal.org/}
It is important to stress that this is still a local interaction rule, albeit in the topological
rather than metric sense.

These choices are motivated by experimental evidence, gathered in
starling flocks \cite{Ballerini2008} and in other social vertebrates
\cite{Gautrais}, that individual do
not interact with neighbours chosen inside a certain fixed range, but
rather with a more or less fixed number of neighbours regardless of
local density. One can think that visual perception, limited by
occlusions to the first shell of neighbours, is better modeled by
topological rather than metric interactions.

In topological models, fluctuations in local density do not affect
the interaction frequency or the number of interacting neighbours, so
that there is no positive feedback on the efficiency of the alignment
interaction. In the absence of an interaction range, it is indeed
possible to rescale lengths in order to always have a unit total
density, $\rho_0=1$. 

Moreover, the long-wavelength instability which
destabilizes the homogeneous ordered phase at the onset of order is
not present in models with topological interactions, and phase
separation is removed \cite{Peshkov}.
The corresponding phase
diagram is much simpler and independent from total density. As the
noise is lowered, one directly crosses from disorder to an
homogeneously ordered, collectively moving, phase. In the absence of
phase separation, the transition is a second order, continuous one,
characterized by a novel set of critical exponents.

\subsection{Long range order in d=2}
\label{MW}

We now turn our attention to the homogeneously ordered phase.

One interesting and, to a certain extent, surprising property emerging
from numerical simulations of the
Vicsek models, is its ability to display true collective motion in
$d=2$, that is, to have a true long range ordered phase in which the
order parameter $\langle \varphi \rangle_t$ is finite for any system
size. This is in apparent contradiction with a well-known theorem due
to Mermin and Wagner (MW) \cite{MW}, stating that no system breaking a continuous
symmetry in two spatial dimensions may achieve long range order (LRO). A
classical example of this theorem is given by the XY model in
$d=2$. In this case, the system may only achieve a lesser kind of order,
called {\it quasi long range order} (QLRO), where the order parameter decays
algebraically with the number of spins $N$, albeit with a very small
exponent \footnote{This exponent
  depends on the equilibrium temperature $T$ and decreases
  monotonously from $1/16$ at the KT transition to $0$ at $t=0$.}. 
While this means that, 
strictly speaking, no order is present asymptotically, a trace of
order can still be found in the algebraically decaying spin-spin
correlation function, and it is thus possible to formally define a phase
transition -- the Kosterlitz-Thouless (KT) transition \cite{KT}.

There is of course a caveat, since the Mermin-Wagner theorem only
applyes to equilibrium systems, and the Vicsek model is of course
out-of-equilibrium. It is however interesting to understand why the
ability of Vicsek particles to move can beat the MW theorem. It is
instructive to consider a simplified argument -- first introduced by John
Toner in his lecture notes on flocking -- \cite{JT} showing that the VM
does so thanks to more efficient information transfer mechanisms. 

First consider XY spins on a $d$ dimensional lattice. Suppose they all
point in the same direction $\langle {\bf s} \rangle$, with the
exception of a single ``mistaken'' spin,
that lies at an angle $\delta \theta_0$ from all the others. How this
mistake will evolve in time on our lattice? Ferromagnetic alignment
cannot simply reset $\delta \theta_0$ to zero: all it can do is to
``iron it out'', spreading it to nearby lattice sites. 
On a lattice, this propagation mechanism is purely diffusive, $\partial_t \delta \theta \sim
\nabla^2 \delta \theta$, and in a time $\tau$ the original error will spread out over a
distance $r \sim \sqrt{\tau}$, or a volume $V_e \sim \tau^{d/2}$. Since
the total error inside the volume is conserved, the error per spin
decays as $\delta \theta \sim \delta\theta_0/V_e \sim \delta\theta_0
\, \tau^{-d/2}$.
This is what happens to a single mistake. However, noise fluctuations
constantly produces local errors with a number of errors per spin 
proportional to time. In a propagation volume $V_e$ one has
$n_d \sim \tau\, V_e \sim \tau^{1+d/2}$ errors. Their combined root mean square,
according to central limit theorem, is $\Omega_e \sim \sqrt{n_e}
\sim \sqrt{\tau\,V_e}$. We are finally in the position to compute the
total error amplitude per spin, that is
\begin{equation}
\Delta \theta \sim \Omega_e / V_e \sim \sqrt{\tau / V_e} 
\sim r^{1-d/2} \to \left\{
\begin{array}{l r}
0 & d>2 \\
\infty & d<2 \\
\ln L \to \infty & d=2
\end{array}\right.
\label{eq:MW}
\end{equation}
If $d>2$, Eq. (\ref{eq:MW}) predicts that fluctuation errors per spin
should decay algebraically in space. This means that order is resistant
to fluctuations, and the system displays long range order. On the
other hand, if $d<2$, fluctuations grows algebraically in space, so that
no global order is possible. The case $d=2$ is marginal, with a zero algebraic
exponent but a logarithmic divergence in the system size $L$
\footnote{The logarithmic divergence basically emerges summing up
  contribution over the entire volume; being careful, one has to
  perform integrals such as $\Delta \theta \sim \int_{L^d} dr^d r^{-d} \sim \ln L$.}.
In this case, fluctuations are still unbounded, but only
logarithmically, so that the order is destroyed extremely slowly
and the equilibrium system displays QLRO. Note that the fact that we are breaking a
{\it continuous} symmetry is essential to this argument. Only in this
case, in fact, arbitrary small fluctuations can induce an arbitrarily small mistake 
$\delta \theta_0$ in spin orientation.

In the VM, however, orientation fluctuations are coupled to
motion. Indeed, fluctuations induce a separation between particles of order 
$\delta x_\perp  \sim v_0 \tau \sin \Delta \theta \sim \tau \Delta
\theta $ in the directions transversal to the mean  direction of motion, and $\delta x_{\parallel} \sim v_0 \tau (1-\cos \Delta \theta) \sim \tau \Delta
\theta^2 $ in the longitudinal direction,
so that two different mechanisms compete to transport orientation information:
particle motion and standard diffusion. The propagation volume is
readily decomposed in its transversal and longitudinal directions $V_e
\sim w_\perp^{d-1}\, w_{\parallel}$ (see Fig.~\ref{fig:3}a), where we have
\begin{eqnarray}
w_\perp &\sim& \delta x_\perp +D_\perp \tau^{1/2} \sim \tau \Delta
\theta +D_\perp \tau^{1/2} \,\dot{\sim}\, \tau^{\gamma_{\perp}} \label{eq:w1}\\
 w_{\parallel} &\sim& \delta x_{\parallel} +D_{\parallel}  \tau^{1/2} \sim \tau \Delta
 \theta^2  +D_{\parallel} \tau^{1/2} \,\dot{\sim}\, \tau^{\gamma_{\parallel}}
\label{eq:w2}
\end{eqnarray}
so that the error per spin in the Vicsek model is given by 
\begin{equation}
\Delta \theta \sim \frac{\tau^{1/2}}{\sqrt{w_\perp^{d-1}\, w_{\parallel}}}
\,\dot{\sim}\, \tau^\gamma
\label{eq:theta}  
\end{equation}
The three equations (\ref{eq:w1})-(\ref{eq:theta}), where we have
introduced the three unknown exponents $\gamma$, $\gamma_{\perp}$ and
$\gamma_{\parallel}$, should be solved simultaneously. They yield a system
of three linear equations in the three unknown exponents 
\begin{eqnarray}
2 \gamma = 1 - \gamma_{\parallel}-(d-1)\gamma_\perp \nonumber\\
\gamma_{\perp}=\mbox{max}\left(1+\gamma, \,\frac{1}{2}\right)\\
\gamma_{\parallel}=\mbox{max}\left(1+2\gamma, \,\frac{1}{2}\right) \nonumber
\label{pip}
\end{eqnarray}
which can be
readily solved. The explicit solution depends on the dimension
$d$. Three different cases are in order. For $d\geq 4$ one has
\begin{equation}
\gamma=\frac{1}{2}-\frac{d}{4}\;\;\mbox{and}\;\; \gamma_{\perp}=\gamma_{\parallel}=\frac{1}{2}
\label{pip1}
\end{equation}
so that above the {\it upper critical dimension} $d_c=4$ the sistem is
fully diffusive and $\gamma <0$. For $7/3\leq d < 4$ transversal
propagation is  superdiffusive and we have
\begin{equation}
\gamma=\frac{3-2d}{2(d+1)}\;,\;\;\gamma_{\perp}=\frac{5}{2(d+1)} \;\;\mbox{and}\;\; \gamma_{\parallel}=\frac{1}{2}
\label{pip2}
\end{equation}
with again a negative $\gamma$. Finally, for $d<7/3$ our simple
argument also predicts superdiffusion propagation also in the
longitudinal direction:
\begin{equation}
\gamma=\frac{1-d}{d+3}\;,\;\;\gamma_{\perp}=\frac{4}{d+3} \;\;\mbox{and}\;\; \gamma_{\parallel}=\frac{5-d}{d+3}
\label{pip3}
\end{equation}
which gives $\gamma<0$ for any $d>1$, so that
orientation fluctuations are suppressed on large scales and the VM can attain long ranger order in any $d>1$, thanks to the non-equilibrium,
self propelled nature of its particles\footnote{Note that this is a
  self consitency argument. Eqs. (\ref{eq:w1})-(\ref{eq:w2}) only
  holds if the system shows LRO, and therefore are invalid for $d\leq1$.} The fact that, below the upper
critical dimension $d_c=4$, particle motion dominates over
simple diffusion -- resulting in a superdiffusive propagation -- is related to the so called {\it breakdown of
    linearized hydrodynamics}. This phenomenon can be studied more
  rigorously by a dynamical renormalization group (DRG) study of the hydrodynamic
  equations for the Vicsek universality class, first obtaines by Toner
  \& Tu by by symmetry arguments \cite{TT1, TT2, TT3, Ramaswamy}. Their detailed analysis
  clearly lies out of the scope of this notes, but it is worth
  mentioning that DRG calculations suggest that it is only in the transversal
  direction that particle motion dominates over
simple diffusion. This consideration forces $\gamma_{\parallel}=1/2$ in the above
argument. This invalidates Eq. (\ref{pip3}) and extends Eq. (\ref{pip2})
below $d=7/3$, yelding $\gamma<0$ and thus LRO in any dimension
larger than $d=3/2$.

Finally, we also note that generically $w_{\perp} >> w_{\parallel}$ so that
fluctuations propagate much slower in the longitudinal directions than
in the transversal ones (see Fig.~\ref{fig:3}a). This spatial anisotropy is of course due to
the symmetry breaking process. Once a direction of motion is picked up,
spatial isotropy is broken and the longitudinal direction can have different
scaling properties from the transversal ones.


\subsection{The Toner \& Tu phase: scale free correlations and anomalous density fluctuations}

The homogeneous ordered phase of the Vicsek class is sometimes referred to as the 
{\it Toner \& Tu phase}, after the authors of the pioneering papers
that first discussed its hydrodynamic behavior \cite{TT1, TT2, TT3}. In this section we
briefly discuss its most important properties, which hold for both
metric and topological interactions.

It is well known that in systems where a continuous symmetry is
spontaneously broken, the entire ordered phase is characterized by
an algebraic decay of its connected correlation functions (i.e. the
corresponding fluctuations correlation function)\cite{Forster, PataPok}. This is also true for the
Vicsek model; moreover, by virtue of the coupling between orientation
and local particle density,
both the density-density and the orientation-orientation connected correlation
functions show an algebraic decay. 
In particular, it is instructive to consider orientation fluctuations 
$\delta {\bf s}_i = {\bf s}_i - \frac{1}{N} \sum_i {\bf s}_i$.
Their equal time, two points correlation function is defined as
\begin{equation}
C_s(r) =\left\langle \frac{\sum_{ij} \delta {\bf s}_i \cdot \delta {\bf s}_j \; \delta(r-r_{ij})}{\sum_{ij}
  \delta(r-r_{ij})} \right\rangle \ ,
\label{corr}
\end{equation}
where $r_{ij}$ is the distance between particle $i$ and $j$ and $\langle \cdot \rangle$ is an
average over realizations (or time in a stationary states).
It can be shown that one has $C_s(r) \sim r^{-\chi}$. 

\begin{figure*}[t]
\centering
\includegraphics[width=0.56\textwidth]{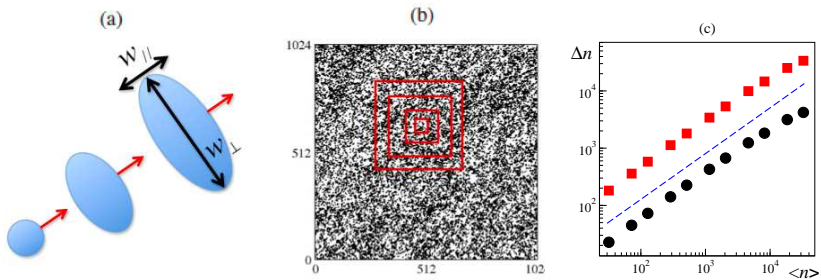}
\includegraphics[width=0.24\textwidth]{Fig3C.eps}
\caption{(a) Schematic representation of information propagation in
  $d=2$. (b) Computing the mean number of particle $\langle n \rangle$
  and its root mean square fluctuations $\Delta n$ in boxes of
  different linear size $\ell$, one can numerically explore the relation $\Delta n
\sim \langle n \rangle^\alpha$. (c) Log-log plot of the numerical results for the VM in
$d=2$ (red squares) amd $d=3$ (black dots). The dashed blue line
corresponds to an exponent $\alpha=0.8$. For more details, see reference \cite{Ginelli2008}.}
\label{fig:3}
\end{figure*}

In systems of finite linear size $L$, due to the global constraint $\sum_{i}
\delta {\bf s}_{i}=0$, the correlation function has a zero,
which can be used as a finite-size definition of the correlation
length $\xi$, $C_s(r=\xi)=0$. As a consequence of the spontaneous
symmetry breaking one has $\xi \sim L$, i.e. the correlation length
scales with the system size. In finite systems one can thus write
\begin{equation}
C_s(r,L) = r^{-\chi}\,g\left(\frac{r}{\xi}\right)
\end{equation}
where $g(r)$ is a universal scaling function with $g(1)=0$. 

We have just shown that in the Vicsek class
orientation (or velocity) fluctuations are {\it scale free}. While a
rigorous demonstration is beyond the scope of these notes,
it is important to remark that this is just a consequence of the
spontaneous breaking of a continuous symmetry. The concept of scale free
correlations in collective motion, in fact, received a certain
attention after they have been measured in starling flocks observed
in the wild \cite{Cavagna2010}.

The algebraic nature of correlation functions has a number of other
non-trivial consequences. The so called giant particles number 
fluctuations are one of the most relevant. We begin giving an
operative, computational definition. Define a box of linear size
$\ell$ inside your system, containing $n_t$ particles at time $t$. One can then measure the mean number of
particles $\langle n \rangle$ contained in the box by taking a mean in time over different
countings. In the homogeneous phase, this will be simply given
by $\langle n \rangle = \rho_0\, \ell^d $. Together with the mean, one can also measure root
mean square fluctuations $\Delta n = \left(\langle (n_t - \langle n
\rangle)^2 \rangle \right)^{1/2}$. 
By considering boxes of different size $\ell$ (see
Fig.~\ref{fig:3}b), one can then explore numerically the relation
between the mean and its fluctuations 
\begin{equation}
\Delta n
\sim \langle n \rangle^\alpha
\label{GDF}
\end{equation}
In equilibrium systems, away from critical point one, has generally
$\alpha=1/2$ in agreement with the central limit theorem, but numerical
simulations \cite{Ginelli2008} show that in the entire Toner \& Tu phase one has $\alpha \approx
0.8$ in both two and three spatial dimensions, as shown in
Fig.~\ref{fig:3}c. Fluctuations in number density are anomalously
large in the Vicsek class!

This is indeed another manifestation of the slow power-law decay of
correlations. A slow enough decay in space of local density\footnote{In numerical experiments, local
  density $\rho({\bf r}, t) = n_t /\ell^d$ can be measured through a
  suitable space coarse-graining over a volume $\ell^d$.} fluctuations
$\delta \rho({\bf r}, t)$ correlations, corresponds indeed to  
an algebric divergence behavior at small frequencies ${\bf q}$ in Fourier space
\begin{equation}
S({\bf q}, t) \equiv \langle \delta \hat{\rho} ({\bf q}, t) \delta \hat{\rho} ({\bf q}, t)
    \rangle \sim \frac{1}{q^\sigma}\;\;\;\;\mbox{for}\;q \to 0
\end{equation}
(where $q=|{\bf q}|$), as opposed to ordinary equilibrium systems where $\sigma=0$.
The small frequency behavior of the stationary density structure factor $S({\bf q})$
gives indeed the fluctuations to mean ratio in the limit of a large
particle number,
\begin{equation}
S({\bf q}\to 0) = \left [ \frac{\Delta n^2}{\langle n
    \rangle}\right]_{n \to \infty}
\end{equation}
Remembering that $\langle n \rangle = \ell^d \rho_0$, and that
transforming back into Fourier space one has $S({\bf q}\to 0) \sim
\frac{1}{q^\sigma} \sim \ell^\sigma$ (the box linear extension $\ell$ being a
small frequency cutoff), one obtains \cite{Rama-GDF}
\begin{equation}
\Delta n
\sim \langle n \rangle^{1/2 + \sigma/(2 d)}
\label{GDF2}
\end{equation}
or 
\begin{equation}
\alpha=\frac{1}{2} + \frac{\sigma}{2 d}
\label{alpha}
\end{equation} 
As anticipated, the equilibrium result
$\alpha=1/2$ is recovered when the structure factor is finite for
$q\to 0$, that is for $\sigma=0$.

The argument given above is slightly simplified in implicitly assuming spatial isotropy of
correlation functions and of the corresponding structure factor. We
indeed know that this is not the case: due to symmetry breaking
spatial isotropy is broken, and correlation functions show different
algebraic behaviors in the transversal and longitudinal directions. In
fact, by measuring means and fluctuations in square boxes, we are
taking an average over the different directions. Correspondingly, in
the above argument, one should average $S({\bf q}\to 0)$ over all
directions. In the Appendix we carry on this procedure in detail making
use of Toner \& Tu theory predictions for $S({\bf q})$ \cite{TT2}. It
yields an estimate of $\alpha=4/5$ for $d=2$ and $\alpha=23/30$ for
$d=3$. Note that the estimates for $d=2$ and $d=3$ are very close to
each other and in substantial agreement with current numerical data as
shown in Fig.~\ref{fig:3}c. 

\subsection{Models with attractive/repulsive interactions and surface tension}

It is finally worth noticing that the alignment rule alone is not able
to maintain the cohesion a of a finite flock in open
space. Fluctuations, in fact, will inevitably pull apart particles one
from each other, finally disintegrating the flock. As already
mentioned, in numerical simulations this problem is usually solved by introducing
periodic boundary conditions, an appropriate choice when one is
interested in the bulk, asymptotic properties of the Vicsek
class. However, if one wants to simulate a finite group in open space,
some attractive interaction should be added to introduce a surface
tension and stabilize the finite flock. Attraction (together with
short range repulsion) was already present in Ref. \cite{Reynolds},
where a pioneering flocking model has been proposed in the context of computer graphics,
but the first study of a VM model with cohesion in a statistical
physics context has been performed in Ref. \cite{Gregoire2003}, where
Eq. (\ref{eq:2bis}) has been modified by adding an
attraction/repulsion term. One has
\begin{equation}
{\bf s}_i^{t + \Delta t} = \frac{\sum_j n_{ij}^t {\bf s}_j^t + \beta
  \sum_j n_{ij}^t f(r_{ij}) \,{\bf e}_{ij}+ \eta \,
  m_i {\bm \xi}_i^t}{\left\| \sum_j n_{ij}^t {\bf s}_j^t + \beta
  \sum_j n_{ij}^t f(r_{ij}) \,{\bf e}_{ij} + \eta \,
  m_i {\bm \xi}_i^t \right\|}
\label{eq:att}
\end{equation} 
where ${\bf e}_{ij}$ is the unit vector going from particle $i$ to $j$
$r_{ij}=|{\bf r}_i - {\bf r}_j|$ is the reciprocal distance and in
\cite{Gregoire2003} topological interaction were used. Here $f$ is a two body force,
repulsive at short range and attracting further away. For instance,
one can chose $f(r)=\mbox{min}(1, r-r_e)$, where $r_e$ is the equilibrium
distance. By increasing the cohesion parameter $\beta$, it has been shown that
the finite flock can pass from a gas phase -- where the group
disintegrates in open space -- to a (moving) liquid one and
eventually to a (moving) crystal phase. The effect of strong repulsion alone added to alignment has been
discussed in \cite{Farrel}.\\

\section{Concluding remarks}

In these notes, we have discussed the Vicsek model and its relative
``universality class'' by making use of numerical experiments and of a
number of illustrative but somehow simplified arguments. A more
rigorous analytical treatment of the VM asymptotic properties is
given by hydrodynamic theories, but their detailed discussions clearly
lies out of the scope of this lecture notes. The interested reader should
consult the original work of Toner \& Tu on phenomenological
hydrodynamics \cite{TT1, TT2, TT3, Ramaswamy}, where an RG approach to the study
of the homogeneous ordered phase
is carried on, and the Boltzmann-Ginzburg-Landau approach developed in \cite{Bertin1, Bertin2}.\\

While the Vicsek universality class is robust to many variations, such
as changes in the way the noise is implementes (as long as no
long-range correlations are introduced) or the details of the local
alignment interaction (but relevant changes can be introduced switching the
interaction from metric to topological as discussed in Section
\ref{topo}), changes in some fundamental features are typically relevant. 
Modifying the nature of broken symmetry, for instance, is a typical
example of such a change. 
For example, one may consider nematic rather than
ferromagnetic alignment, without altering the polar
self-propelled nature of particles (the so called self propelled rods model)
\cite{Ginelli2010a}, or consider altogether completely nematic
particles (which have a preferred axis of motion but not a well defined
direction) such as in active nematics \cite{Ginelli2006}. These models are
relevant to the modelling of elongated active particles interacting by
volume exclusion forces, which typically induce an effective nematic interaction.
In general, these so called {\it Vicsek-like} models constitute different
universality classes, but share a very similar phase diagram
structure with the Vicsek class: the phase diagram of all metric
Vicsek-like models, for isntance, exhibit a phase separated regime (possibly with different
symmetries/properties w.r.t. the VM) taking place at the onset of
order and separating the disordered from the homogeneously ordered phase.  

Other relevant changes include violation of particles number
conservation, as discussed in Ref. \cite{MF}, or -- as previously
discussed -- the inclusion of
momentum conservation and long-ranged hydrodynamic interactions. 

\subsection*{Acknowledgments}
I acknowledge support from the Marie Curie Career Integration Grant
(CIG) PCIG13-GA-2013-618399. I am also indebited with H. Chat\'e, J. Toner
and S. Ramaswamy for many lectures and spirited discussions that found their
way into these notes.  

\clearpage
\appendix

\section{Appendix -- The anomalous density fluctuations exponent}

In this appendix, we compute explicitly the anomalous density fluctuations
exponent making use of the results of Toner \& Tu theory. The density
structure factor has an anysotropic structure and it is given by (in
units of the interaction distance $R_0$)
\cite{TT2}
\begin{equation}
S({\bf q}) \sim \left\{\begin{array}{l}
q_{\perp}^{1-d-\zeta-2\chi}\;\;,\;\;\;\;\;\;\;\;\;\;\;\;  q_{\parallel}\ll q_{\perp}\\
q_{ \parallel}^{-2}q_{\perp}^{3-d-\zeta-2\chi}\;\;,\;\;\;\;\;\;\;q_{\perp}^\zeta \gg q_{\parallel} \gg q_{\perp}\\
q_{\parallel}^{-3+(1-d-2\chi)/\zeta}q_{\perp}^2\;,\;\;\;\;q_{\perp}^\zeta \ll q_{\parallel}
\end{array}\right.
\label{Sq1}
\end{equation}
where $q_{\parallel}$ and $q_{\perp}$ are (respectively) the
projection of the Fourier space vector ${\bf q}$ in the
longitudinal and transversal directions w.r.t. the direction of
motion. The two exponents $\zeta$ and $\chi$ are scaling exponents
for which the DRG flows to a fixed point. 
According to a conjecture first put forward in \cite{TT2}, in
any dimension $3/2 \! \leq \! d \! \leq \! 4$ they are 
\begin{equation}
\chi= {3 - 2 d \over 5}\;,\;\;\; 
\zeta= {d + 1 \over 5}.
\label{scaling-exp}
\end{equation}
While this conjecture has
never been proven rigorously, there is a reasonable numerical \cite{TT4, Ginelli2008} evidence supporting the
above scaling exponent values  for  $d \!=\! 2$ and, to a lesser
extent, $d \!=\! 3$. In the following we will assume the above values
hold.

\begin{figure}[t]
\centering
\includegraphics[width=0.4\textwidth]{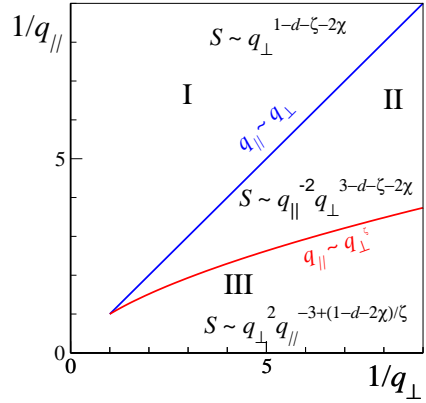}
\caption{Graphical representation of density structure factor scaling.}
\label{figA}
\end{figure}

We can visualize the three different sectors which in Eq. (\ref{Sq1})
determines the scaling of the density structure factor as in
Fig.~\ref{figA}. In particular, we are interested in the scaling
behavior as one approaches ${\bf q}=0$ along different paths in the
$(q_{\perp},q_{\parallel})$ plane (or moves towards infinity in
the real axis representation $(1/q_{\perp},1/q_{\parallel})$. 
It is easy to see that moving towards infinity along the line
$q_\parallel \sim q_\perp \sim q$ the structure factor picks up a divergence  
\begin{equation}
S({\bf q}) \sim q^{1-d-\zeta-2\chi} \sim q^{-2(d+1)/5}
\label{max}
\end{equation}
This is actually the strongest possible divergence in any $d < 4$. Moving to infinity
along the line $q_\parallel \sim q_\perp^\zeta $, for instance, gives
\begin{equation}
S({\bf q}) \sim q_\perp^{3-d-3\zeta-2\chi} \sim q_\perp^{(6-4d)/5}
\label{max2}
\end{equation}
while chosing other paths towards ${\bf q}=0$ lying in the sector I,
II or III Fig.~\ref{figA} also produces weaker divergences or no
divergences at all. This can be checked by chosing a family of paths
$q_\parallel \sim q_\perp^\nu $. The value of the exponent $\nu$
determines the chosen sector for our path, with $\nu>1$ corresponding
to sector I, $1>\nu>\zeta$ to sector II and $\nu<\zeta$ to sector
III. To summarize, the structure factor is dominated by divergences
along the $q_\parallel \sim q_\perp $ line,
\begin{equation}
S({\bf q}) \sim q^{-\sigma}     \;\;\;\;\mbox{with}\;\; \sigma = \frac{2}{5}(d+1)
\label{max3}
\end{equation}
By Eqs. (\ref{alpha}) this finally gives
\begin{equation}
\alpha = \frac{1}{2}+\frac{d+1}{5 d}
\end{equation} 





\begin{thebibliography}{99}

\bibitem{Parrish} {\it Animal Groups in Three Dimensions}, edited by J. K. Parrish
and W. M. Hamner (Cambridge University Press, Cambridge,
U.K., 1997)..

\bibitem{Ginelli2015} F. Ginelli {\it et al.}
  Proc. Natl. Acad. Sci. USA {\bf 112}, 12729 (2015).

\bibitem{Swinney2010} 
H.P. Zhang {\it et al.}, 
Proc. Natl. Acad. Sci. USA {\bf 107}, 13626 (2010).

\bibitem{Trepat} X. Trepat {\it et al.}, Nature Physics {bs 5} 426 (2009).

\bibitem{Schaller2010}
V. Schaller {\it et al.}, Nature {\bf 467}, 73 (2010).

\bibitem{Pliny} Pliny, {\it Natural History}, translated by H. Rackham (Harvard
University Press, 1968), Vol. III, book 10, xxxii, p. 63

\bibitem{Ramaswamy2010} S. Ramaswamy, Annu. Rev. Condens. Matter
  Phys. {\bf 1}, 323 (2010).

\bibitem{Palacci} J. Palacci, S. Sacanna, A. Preska Steinberg,
  D.J. Pine, P.M. Chaikin, Science {\bf 339} 936 (2013).

\bibitem{Narayan}
V Narayan, S Ramaswamy, N Menon, Science {\bf 317}, 105 (2007).

\bibitem{Desaigne2010} J. Desaigne, O. Dauchot, and H. Chat\'e, Phys. Rev. Lett. {\bf 105}, 098001 (2010). 

\bibitem{Forster} D. Forster, {\it Hydrodynamic Fluctuations, Broken
    Symmetry and Correlation Functions}, Frontiers in Physics {\bf 47}, XIX,
  326 S., (W. A. Benjamin, Inc, 1975).




\bibitem{Vicsek95} T. Vicsek {\it et al.}, 
Phys. Rev. Lett. {\bf 75}, 1226 (1995). 

\bibitem{Gregoire2004} 
G.~Gr\'egoire and H.~Chat\'e, Phys. Rev. Lett. {\bf 92}, 025702 (2004);

\bibitem{Ginelli2008}
H. Chat\'e, {\it et al.},  Phys. Rev. E {\bf 77}, 046113 (2008).

\bibitem{Ginelli07}
H. Chat\'e, F. Ginelli and G.~Gr\'egoire, Phys. Rev. Lett. {\bf 99},, 229601 (2007).

\bibitem{Bertin1} E. Bertin, M. Droz, and G. Gr\'egoire, Phys. Rev. E
  {\bf 74}, 022101 (2006). 

\bibitem{Bertin2} E. Bertin, M. Droz, and G. Gr\'egoire, J. Phys. A {\bf 42}, 445001 (2009).

\bibitem{Tailleur2}
A P. Solon,  H. Chat\'e, J. Tailleur Phys. Rev. Lett. {\bf 114}, 068101 (2015); 


\bibitem{Ginelli2010b}
F. Ginelli, H. Chat\'e, Phys. Rev. Lett. {\bf 105}, 168103 (2010)

\bibitem{Ballerini2008} 
M. Ballerini {\it et al.} Proc. Natl. Acad. Sci. USA {\bf 105},1232 (2008).

\bibitem{Gautrais}
J. Gautrais {\it et al.} PLoS Comp. Bio. {\bf 8} e1002678 (2012).

\bibitem{Peshkov} A. Peshkov {\it et al.}, Phys. Rev. Lett. {\bf 109}, 098101 (2012).

\bibitem{MW} N. D. Mermin and H. Wagner, Phys. Rev. Lett. {\bf 17}, 1133
(1966). 


\bibitem{KT} J. M. Kosterlitz and D. Thouless, J. Phys. C {\bf 6}, 1181
(1973).

\bibitem{JT} J. Toner, {\it unpublished}. I am deeply indebited to
  John Toner for having introduced me to this simple but illuminating argument.

\bibitem{TT1}
J. Toner and Y. Tu, Phys. Rev. Lett. {\bf 75}, 4326 (1995). 

\bibitem{TT2}
J. Toner and Y. Tu, Phys. Rev. E {\bf 58}, 4828 (1998). 

\bibitem{TT3} J. Toner, Phys. Rev. E {\bf 86}, 031918 (2012).

\bibitem{Ramaswamy}
J. Toner, Y. Tu, and S. Ramaswamy, Ann. Phys. (Berlin) {\bf 318}, 170
(2005).

\bibitem{PataPok}
V. L. Pokrovskii and A. Z. Pata{\u s}hinskii, Fluctuation Theory of Phase Transitions, 2nd ed. (Pergamon, Oxford, 1979; Nauka, Moscow, 1982).

\bibitem{Cavagna2010} 
A. Cavagna {\it et al.} Proc. Natl. Acad. Sci. USA {\bf 107},11865 (2010).

\bibitem{Rama-GDF} S. Ramaswamy, R.A. Simha, and J. Toner,
  Europhys. Lett. {\bf 62}, 196 (2003).

\bibitem{Reynolds}
C.W. Reynolds, Computer Graphics, {\bf 21}, 25 (1987). 

\bibitem{Gregoire2003} G. Gr\'egoire, H. Chat\'e and Y. Tu  Physica D
  {\bf 181}, 157 (2003).

\bibitem{Farrel}
F. D. C. Farrell, M. C. Marchetti, D. Marenduzzo, and J. Tailleur
Phys. Rev. Lett. {\bf 108}, 248101 (2012). 

\bibitem{Ginelli2010a}
F. Ginelli, F. Peruani, M. Baer, H. Chat\'e, Phys. Rev. Lett., {\bf 104},
184502 (2010).

\bibitem{Ginelli2006} 
H. Chat\'e, F. Ginelli, R. Montagne, Phys. Rev. Lett., {\bf 96} 180602
(2006).

\bibitem{MF} J. Toner, Phys. Rev. Lett., {\bf 108}, 088102 (2012).

\bibitem{TT4}  Y.-h. Tu, M.\ Ulm and  J.\  Toner, Phys.\ Rev.\ Lett.\  {\bf 80}, 4819 (1998).



















\end{thebibliography}
\end{document}